\documentclass[twocolumn,prb]{revtex4}
\usepackage{amsmath}
\usepackage{amsfonts}
\usepackage{amssymb}
\usepackage{graphics}
\usepackage{float}
\usepackage{graphicx}
\usepackage{epstopdf}
\usepackage[compact]{titlesec}
\usepackage{natbib}
\usepackage{threeparttable}
\usepackage{lipsum}

\begin{document}
\title{Proximity induced ferromagnetism, superconductivity, and finite-size effects on the surface and edge states of topological insulator nanostructures}
\author{Parijat Sengupta$^{1}$}
\email{psengupta@cae.wisc.edu}
\author{Tillmann Kubis$^{2}$}
\author{Yaohua Tan$^{2}$}
\author{Gerhard Klimeck$^{2}$}
\affiliation{$^{1}$Dept of Material Science and Engineering, University of Wisconsin, Madison, WI 53706 \\
$^{2}$Dept of Electrical and Computer Engineering, Purdue University, West
Lafayette, IN, 47907 
}

\begin{abstract}
Bi$_{2}$Te$_{3}$ and Bi$_{2}$Se$_{3}$ are well known 3D-topological insulators. Films made of these materials exhibit metal-like surface states with a Dirac dispersion and possess high mobility. The high mobility metal-like surface states can serve as channel material for TI-based field effect transistors. While such a transistor offers superior terminal characteristics, they suffer from an inherent zero band gap problem. The absence of a band gap for the surface states prevents an easy turn-off mechanism. In this work, techniques that can be employed to easily open a band gap for the TI surface states is introduced. Two approaches are described: 1) Coating the surface states with a ferromagnet which has a controllable magnetization axis. The magnetization strength of the ferromagnet is incorporated as an exchange interaction term in the Hamiltonian. 2) An \textit{s}-wave superconductor, because of the proximity effect, when coupled to a 3D-TI opens a band gap on the surface. This TI-superconductor heterostructure is modeled using the Bogoliubov-de Gennes Hamiltonian. A comparison demonstrating the finite size effects on surface states of a 3D-TI and edge states of a CdTe/HgTe/CdTe-based 2D-TI is also presented. 3D-TI nanostructures can be reduced to dimensions as low as 10.0 $ \mathrm{nm} $ in contrast to 2D-TI structures which require a thickness of at least 100.0 $ \mathrm{nm} $. All calculations are performed using the continuum four-band k.p Hamiltonian.

\end{abstract}
\maketitle

\section{Introduction}
\label{intro} 
Topological insulators (TI) are a new class of materials whose surfaces/edges host bound spin-polarized Dirac fermions with high mobility. These states, in a time reversal invariant system are protected against perturbation and non-magnetic disorder.~\cite{hasanrmp,fuprb07,qi2011topological,kaneqsh05,murakami07} Well-known examples of materials with such properties include Bi$_{2}$Te$_{3}$, Bi$_{2}$Se$_{3}$, Sb$_{2}$Te$_{3}$, Bi$_{x}$Sb$_{1-x}$ alloy, and the CdTe/HgTe/CdTe quantum well heterostructure. Bi$_{2}$Te$_{3}$, Bi$_{2}$Se$_{3}$, Sb$_{2}$Te$_{3}$, and Bi$_{x}$Sb$_{1-x}$ alloy belong to the class of 3D topological insulators (3D TI)~\cite{xia09obsv,chen09exp,roy09} and host bound states on their surface. These states are characterized by a linear Hamiltonian and form a band-gap closing Dirac cone on each surface at the $ \Gamma $ point.

In this work, methods that alter the surface state dispersion through proximity induced ferromagnetism, superconductivity, and finite-size effects are described. Alteration of the surface dispersion is driven in part by the need to utilize these materials as field effect transistors (FETs) that uses the highly mobile zero-gap surface or edge states as the conducting channel. The zero band gap surface states though prohibit an easy turn-off mechanism and therefore inducing a band gap is essential for successful transistor operation. A key idea to open a band gap is destruction of time reversal symmetry~\cite{carpentier2014topology,franz2013topological} which is responsible for protecting the surface states. Time reversal symmetry, which in principle, can be destroyed in the presence of an external perpendicular magnetic field~\cite{mildred2008group} and appears to be a straightforward approach in a standard experimental set-up, may not be applied easily to a miniaturized device on account of electromagnetic interference with the surrounding circuitry. Alternatively, the surface can be coated with a ferromagnet with an out-of-plane magnetization component to replicate this behaviour. 

It is now well established that doping with Cu or Pb can introduce superconducting states in Bi$_{2}$Se$_{3}$. Additionally, through the proximity effect at the interface between a superconductor (SC) and topological insulator, the superconductor's wave functions can penetrate the surface of a topological insulator and induce superconductivity. This induced superconductivity, by virtue of its intrinsic energy gap between the Fermi-level and the superconducting ground state offers a possible way to open a band-gap in a topological insulator. A modified version of the Bogoliubov-de Gennes (BdG) Hamiltonian for a 3D-TI and an \textit{s}-wave superconductor is used to compute the dispersion relationship of a TI-SC heterostructure.

The paper is structured as follows: In Section~\ref{meta}, the 4-band k.p Hamiltonian to compute the surface and edge state energy dispersion is presented. Dispersion relations obtained through them are utilized later to investigate finite-size effects on surface and edge states of 3D and 2D TIs respectively. The 4-band Hamiltonian is modified via an exchange-interaction term to model proximity-induced ferromagnetism on the surface of a 3D TI. The TI-SC heterostructure is represented by a BdG Hamiltonian and dispersion when the superconductor is an \textit{s}-wave or \textit{p}-wave type is worked out analytically. Spin matrices using the basis states of the 4-band Hamiltonian are set up to compute spin-polarization of surface states. Numerical calculations based on the Hamiltonians created for several TI systems in Section~\ref{meta} are collected in Section~\ref{res}. The paper concludes by summarizing key results in Section~\ref{conc}. 

\section{Model and Theory}
\label{meta}

Surface and edge states in topological insulators are characterized by a linear dispersion and Dirac cones. They further depend on dimensions and growth conditions of the structure that host them. Low-energy continuum models for 3D and 2D TIs used in deriving results included in this paper are described in this section.

\subsection{Four-band k.p method for 3D topological insulators}
The dispersion relations of Bi$_{2}$Te$_{3}$, Bi$_{2}$Se$_{3}$, and Sb$_{2}$Te$_{3}$ films are computed using a 4-band k.p Hamiltonian. The 4-band Hamiltonian~\cite{zhang09} is constructed (Eq.~\ref{eqn1}) in terms of the four lowest low-lying states $ \vert P1_{z}^{+} \uparrow \rangle $, $ \vert P2_{z}^{-} \uparrow \rangle $, $ \vert P1_{z}^{+} \downarrow \rangle $, and $ \vert P2_{z}^{-} \downarrow \rangle $. Additional warping effects~\cite{fu2009hexagonal} that involve the $k^{3}$ term are omitted in this low-energy effective Hamiltonian.

\begin{align}
\label{eqn1}
H(k) = \epsilon(k) + \begin{pmatrix}
M(k) & A_{1}k_{z} & 0 & A_{2}k_{-} \\
A_{1}k_{z} & -M(k) & A_{2}k_{-} & 0 \\
0 & A_{2}k_{+} & M(k) & -A_{1}k_{z} \\
A_{2}k_{+} & 0 & -A_{1}k_{z} & -M(k) \\
\end{pmatrix}
\end{align}
where $ \epsilon(k) = C + D_{1}k_{z}^{2} + D_{2}k_{\perp}^{2}$, $ M(k) = M_{0} + B_{1}k_{z}^{2} + B_{2}k_{\perp}^{2}$ and $ k_{\pm} = k_{x} \pm ik_{y}$. For Bi$_{2}$Te$_{3}$ and Bi$_{2}$Se$_{3}$, the relevant parameters are summarized in Table.~\ref{table1}.
\begin{table}[htb]
\caption{4-band k.p parameters~\cite{liu2010model} for Bi$_{2}$Te$_{3}$ and Bi$_{2}$Se$_{3}$.}
\centering
\label{table1}
\begin{tabular}{lcc}
\noalign{\smallskip} \hline \hline \noalign{\smallskip}
Parameters & Bi$_{2}$Te$_{3}$ & Bi$_{2}$Se$_{3}$ \\\hline
M$_{0}$ (eV) & 0.30 & 0.28 \\
A$_{1}$ (eV \AA ) & 2.26 & 2.2 \\
A$_{2}$ (eV \AA )  & 2.87 & 4.1 \\
B$_{1}$ (eV \AA$^{2}$)  & 10 & 10 \\
B$_{2}$ (eV \AA$^{2}$)  & 57.38 & 56.6 \\
C (eV) & -0.18 & -0.0068 \\
D$_{1}$ (eV \AA$^{2}$) & 6.55 & 1.3 \\
D$_{2}$ (eV \AA$^{2}$) & 29.68 & 19.6 \\
\noalign{\smallskip} \hline \noalign{\smallskip}
\end{tabular}
\end{table} 

\begin{figure}[h]
\includegraphics[scale= 1]{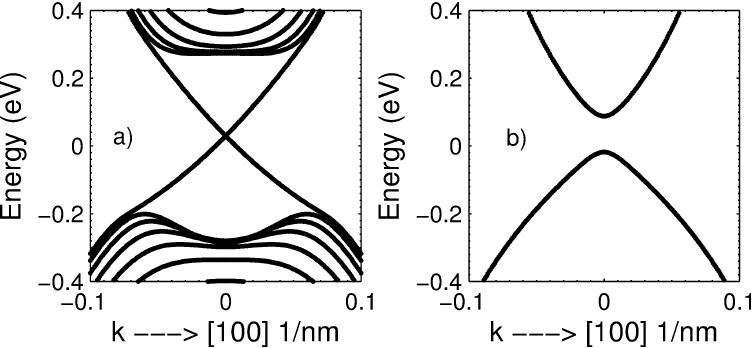}
\caption{Topological insulator surface states (~\ref{fig1}a) around 0.02 $\mathrm{eV}$ for a 20.0 $\mathrm{nm}$ thick (around 20 quintuple layers) Bi$_{2}$Se$_{3}$ film. The dispersion of the thin film (Fig.~\ref{fig1}b) shows two Dirac hyperbolas when the surface states hybridize.}
\label{fig1}
\end{figure}

The dispersion for a 20.0 $\mathrm{nm}$ Bi$_{2}$Se$_{3}$ thick film which is approximately twenty quintuple-layers is shown in Fig.~\ref{fig1}a. The Dirac cone is formed at an energy equal to 0.029 $\mathrm{eV}$ confirming that it is indeed a mid-gap state. The bulk band-gap of Bi$_{2}$Se$_{3}$ is approximately 0.32 $\mathrm{eV}$ at the $\Gamma$ point. In contrast to the thick-film dispersion, the band profile of a 3.0 $\mathrm{nm}$ (approximately three quintuple-layers) Bi$_{2}$Se$_{3}$ film has a finite band gap. In the case of a thin-film, the two surface states hybridize. The hybridization occurs because each state has a definite localization or penetration length. When the penetration length is comparable to film thickness, the opposite spin-resolved bands of the two surfaces will mix. Since bands with identical quantum numbers cannot cross, a gap(Fig.~\ref{fig1}b) opens up at the $\Gamma$ point and the dispersion changes to Dirac-hyperbolas. It must be stated here, that though a thin film on account of intrinsic hybridization, which is a function of dimension, is not a preferred choice to manipulate the band gap. The band gap control must be performed externally as described in the latter part of this paper.

Two variants of a topological insulator film are considered. The first is a free-standing symmetric thin-film of Bi$_{2}$Se$_{3}$. Two exactly degenerate Dirac cones are formed for a symmetric film. The second thin film considered is assumed to grow on a substrate which induces an asymmetry between the two surfaces. The asymmetry can be explained by assuming that the top surface is exposed to vacuum while the bottom surface is fixed to a substrate. Inversion symmetry therefore does not hold~\cite{sakamoto2010spectroscopic,chang2010growth} for such a TI film. This asymmetry is simulated by adding a small symmetry-breaking potential to the Hamiltonian along the confinement direction. It can also be viewed as a simple way of creating two chemically distinct surfaces. The Dirac cones, one on each surface, are therefore no longer degenerate. They are now split as a function of the applied symmetry-breaking potential and positioned at distinct energies. This splitting is similar to a Rashba-split~\cite{bychkov1984properties} in presence of structural inversion asymmetry (SIA). The asymmetry of the surfaces, in this case, causes the SIA.

When the conditions of ultra-thin film and asymmetry are simultaneously fulfilled, it is observed that the two surface bands offer another instance of Rashba-type splitting. To explain this phenomenon, first a low thickness film is considered. The two degenerate massless Dirac cones in this case would mix and open a finite band-gap as explained above. Each Dirac cone can now be represented as a massive spin-degenerate Dirac hyperbola.~\cite{kuemmeth2009giant} If an additional asymmetry is imposed through a potential, the spin degeneracy of each Dirac hyperbola is broken and the bands split along $ k_{\parallel}$ in opposite directions. The degeneracy is only maintained at the $\Gamma$ point. The conventional Rashba-type splitting follows exactly the same pattern, for example in an asymmetric GaAs quantum well.~\cite{e1997spin}

A hallmark of the surface bands of a 3D-topological insulator is their intrinsic complete spin-polarization and locking of the spin perpendicular to the momentum. Within the framework of the 4-band k.p Hamiltonian, expectation value of the three spin-polarized vectors is computed. The operators for the three spin-polarizations are given by Eq.~\ref{eqn2} and Eq.~\ref{eqn3} in the Pauli representation $\sigma_{i}$ $ \lbrace i = x,y,z \rbrace$.

\begin{align}
S_{x} = \begin{pmatrix}
0 & 0 & 1 & 0 \\
0 & 0 & 0 & 1 \\
1 & 0 & 0 & 0 \\
0 & 1 & 0 & 0  \\
\end{pmatrix} ; \qquad
S_{y} = \begin{pmatrix}
0 & 0 & -i & 0 \\
0 & 0 & 0 & -i \\
i & 0 & 0 & 0 \\
0 & i & 0 & 0  \\ 
\end{pmatrix}  \qquad
\label{eqn2}
\end{align}
\begin{equation}
S_{z} = \begin{pmatrix}
1 & 0 & 0 & 0 \\
0 & 1 & 0 & 0 \\
0 & 0 & -1 & 0 \\
0 & 0 & 0 & -1 
\end{pmatrix}
\label{eqn3} 
\end{equation}

\begin{figure}[h]
\includegraphics[scale=0.9]{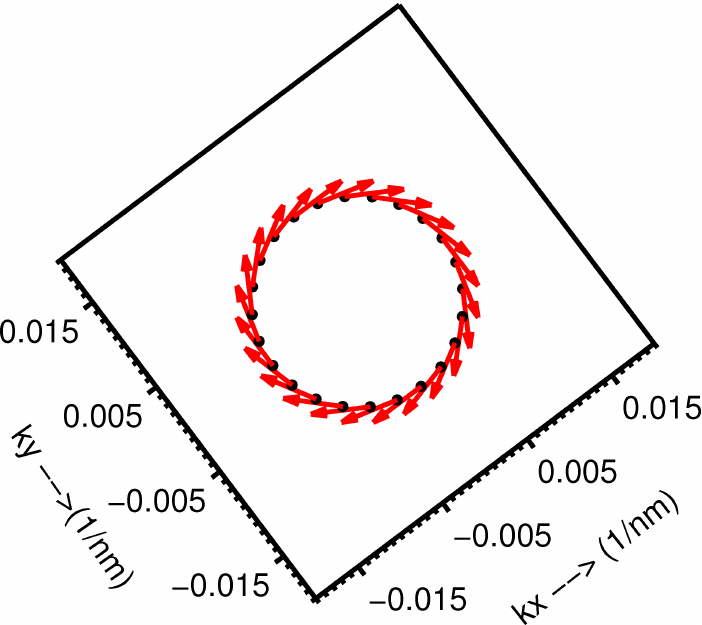}
\caption{The spin polarization confined to the plane in the vicinity of the $\Gamma$ point for a 20.0 $\mathrm{nm} $ thick Bi$_{2}$Se$_{3}$ film. The spin is locked to momentum (which is a radial vector on the circle) shown by the tangential lines on the plot.}
\label{fig2}
\end{figure}

The above matrices are written under the basis set ordered as $ \vert P1_{z}^{+} \uparrow \rangle $, $ \vert P2_{z}^{-} \uparrow \rangle $, $ \vert P1_{z}^{+} \downarrow \rangle $, and $ \vert P2_{z}^{-} \downarrow \rangle $. The expectation value for each spin-polarization operator is calculated in the usual way in Eq.(~\ref{eqn4}).
\begin{equation}
\langle S_{i} \rangle = \int\psi^{*} S_{i} \psi d\tau
\label{eqn4}
\end{equation}
where $ \lbrace i = x,y,z \rbrace $
The spin-polarization shown in Fig.~\ref{fig2} is computed by choosing a $k$-vector of magnitude 0.01 $\mathrm{nm^{-1}}$ at various polar angles.For each such vector, the spin-polarization obtained was perpendicular to momentum with a zero out-of-plane component. The red-arrows in Fig.~\ref{fig2} denote the direction of spin-polarization and is always tangential to the circular contour traced out by the $k$-vector.

\subsection{BHZ Hamiltonian for CdTe/HgTe/CdTe 2D-TI}
2D-TIs which exhibit edge states are also highly mobile and suitable for nanoscale applications.~\cite{sengupta2013design} One example of edge states in a 2D-TI is a CdTe/HgTe/CdTe nanoribbon obtained by quantizing  one of the edges of the corresponding quantum well. It would be useful to compare length scales up to which an edge and surface state can be sustained in a 3D and 2D topological insulator system. While an eight-band k.p Hamiltonian describes the full set of six valence (including spin split-off) and two conduction bands and their mutual interaction through the off-diagonal terms, it is sufficient to focus on bands that exclusively take part in the inversion process~\cite{bernevig2013topological,rothe2010fingerprint}. This interaction of bands is governed by the coupling of conduction and valence states, represented through a linear term as shown in Eq.~\eqref{fullham}.

\begin{align}
\label{fullham}
H(k) = \epsilon(k) + \begin{pmatrix}
M & Ak_{+} & 0 & 0 \\
Ak_{-} & -M & 0 & 0 \\
0 & 0 & M & -Ak_{+} \\
0 & 0 & -Ak_{-} & -M
\end{pmatrix}
\end{align}
where \begin{equation} \epsilon(k) = (C - Dk^{2})I_{4\times4} 
\end{equation}
and \begin{equation}
M = M_{0} - Bk^{2}
\end{equation} $\epsilon(k)$ describes band bending and 2\textit{M$_{0}$} = -\textit{E}$_{g0}$ corresponds to energy gap between bands and is negative in the inverted order bands.

This Hamiltonian is written in the basis of the lowest quantum well sub-bands $\vert E +\rangle $, $\vert H +\rangle $, $\vert E -\rangle$, and $\vert H -\rangle$.
Here $\pm$ stands for the two Kramers partners. The sign of the gap parameter \textit{M} determines if it is a trivial insulator (\textit{M} $>$ 0) or a topological insulator (\textit{M} $<$ 0). Experimentally, \textit{M$_{0}$} is tuned by changing the quantum well width. The parameters \textit{A,B,C,D,and M$_{0}$} are geometry dependent. For numerical calculations performed in this paper, the parameters were set to: $ A = 364.5 meV.nm $, $ B = -686 meV.nm^{2}$, $ C = 0 meV$, $ D = 512 meV.nm^{2} $, and $ M = -10 meV $.  

\subsection{Hamiltonian for 3D-TI and superconductor heterostructure}
 Before a complete Hamiltonian for a TI-superconductor heterostructure can be written, the conventional BCS description~\cite{tinkham2012introduction,de1999superconductivity} of an \textit{s}-wave superconductor must be examined. The BCS Hamiltonian in its simplest form can be written by starting with a Hamiltonian (Eq.~\ref{bcs3}) that describes a many-Fermion system with a spin-independent interaction potential. Such a Hamiltonian in general form is expressed as: 
\begin{equation}
H = \sum_{k\sigma}\varepsilon_{k}a_{k\sigma}^{+}a_{k\sigma} + \dfrac{1}{2\Omega}\sum_{\sigma\sigma^{'}}\sum_{kk^{'}q}\overline{V_{q}}a^{+}_{k+q,\sigma}a^{+}_{k^{'}-q,\sigma^{'}}a_{k\sigma}a_{k^{'}\sigma^{'}}
\label{bcs3}
\end{equation}
By restricting to paired fermions with zero total momentum and opposite spin, the BCS Hamiltonian can be written as
\begin{equation}
H = \sum_{k\sigma}\varepsilon_{k}a_{k\sigma}^{+}a_{k\sigma} + \dfrac{1}{\Omega}\sum_{kk^{'}}V_{k-k^{'}}a^{+}_{k^{'}\uparrow}a^{+}_{-k^{'}\downarrow}a_{k\uparrow}a_{-k\downarrow}
\label{bcs4}
\end{equation}

The spectrum of this Hamiltonian when solved using the Bogoliubov transformation~\cite{fetter2003quantum} yields a band structure~\cite{poole1999handbook} with a gap in the spectrum.  
For studying proximity effect between a superconductor~\cite{orlando1991foundations} and a topological insulator, the 4-band k.p model and the BCS Hamiltonian is used in conjunction. The fundamental assumption of the BCS Hamiltonian is the formation of Cooper pairs which are paired electrons with zero momentum and spin. Superconductivity which is induced on the TI side of the TI-SC heterostructure must therefore agree to this principle. At this point it is worth mentioning again that the four orbitals participating in the electronic bonding process are $ \vert P1_{z}^{+} \uparrow \rangle $, $ \vert P2_{z}^{-} \uparrow \rangle $, $ \vert P1_{z}^{+} \downarrow \rangle $, and $ \vert P2_{z}^{-} \downarrow \rangle $. The composite Hamiltonian for the TI-SC structure similar to the BdG Hamiltonian can now be written as
\begin{equation}
H_{TS} = \left( \begin{array}{cc}
H_{T} - \mu & \Delta  \\
\Delta^{*} & \mu - TH_{T}T^{-1}  \\
\end{array} \right)
\label{bdgprimer}
\end{equation}

In Eq.~\ref{bdgprimer}, $ \mu $ denotes the chemical potential and \emph{T} is the time reversal operator. H$_{TS}$ is the composite Hamiltonian and H$_{T}$ represents the intrinsic 4-band k.p Hamiltonian. In the composite Hamiltonian, Eq.~\ref{bdgprimer}, $ \Delta = \lambda\left(\overrightarrow{r}\right)F\left(\overrightarrow{r}\right)$ is the pair potential where $ F_{\sigma \sigma^{'}}\left(\overrightarrow{r}\right) =  \langle\Psi_{\sigma}\left(\overrightarrow{r}\right)\Psi_{\sigma^{'}}\left(\overrightarrow{r}\right)\rangle $ is the order parameter and $ \lambda\left(\overrightarrow{r}\right) $ characterizes the strength of the attractive interaction potential as established in the BCS formulation. For the \textit{s}-wave superconductor considered here, the pair-potential, $ F_{\sigma \sigma^{'}}\left(\overrightarrow{r}\right) = i\Delta_{0}\sigma_{y} $ is isotropic and just a number. The analytic representation of pair-potential changes to a $ \overrightarrow{k}$ dependent quantity if \emph{p} or \emph{d}-type superconductors are considered. The pair potential $ F_{\sigma \sigma^{'}}\left(\overrightarrow{r}\right)$ denotes a correlated pair of electrons with opposite momentum and spin. For the case of a TI, which is turned in to a superconductor, the orbitals with opposite spin and momentum are paired. The two sets of orbitals in the 4-band TI Hamiltonian are therefore coupled by two pair potentials. The full TI-SC Hamiltonian \emph{H$_{TS}$} in the basis set {$ \vert P1_{z}^{+} \uparrow \rangle $, $ \vert P2_{z}^{-} \uparrow \rangle $, $ \vert P1_{z}^{+} \downarrow \rangle $, $ \vert P2_{z}^{-} \downarrow \rangle $, $ -\vert P1_{z}^{+} \uparrow \rangle $, $ -\vert P2_{z}^{-} \uparrow \rangle $, $ -\vert P1_{z}^{+} \downarrow \rangle $, and $ -\vert P2_{z}^{-} \downarrow \rangle $ } can be now written as
\begin{widetext}
\[
 H_{TS} =  \left( \begin{array}{cccccccc}
\epsilon+ M & A_{1}k_{z} & 0 & A_{2}k_{-} & 0 & 0 & \Delta_{1} & 0 \\
A_{1}k_{z} & \epsilon- M & A_{2}k_{-} & 0 & 0 & 0 & 0 & \Delta_{2} \\
0 & A_{2}k_{+} & \epsilon+ M & -A_{1}k_{z} & -\Delta_{1} & 0 & 0 & 0 \\
A_{2}k_{+} & 0 & -A_{1}k_{z} & \epsilon- M & 0 & -\Delta_{2} & 0 & 0 \\
0 & 0 & -\Delta_{1}^{*} & 0 & -\epsilon- M & A_{1}k_{z} & 0 & A_{2}k_{-} \\
0 & 0 & 0 & -\Delta_{2}^{*} & A_{1}k_{z} & -\epsilon+ M & A_{2}k_{-} & 0 \\
\Delta_{1}^{*} & 0 & 0 & 0 & 0 & A_{2}k_{+} & -\epsilon- M & -A_{1}k_{z} \\
0 & \Delta_{2}^{*} & 0 & 0 & A_{2}k_{+} & 0 & -A_{1}k_{z} & -\epsilon+ M \\
 \end{array} \right) - \mu I_{8 \times 8}
\label{bdg_full}
\]
\end{widetext}
where $ \epsilon(k) = C + D_{1}k_{z}^{2} + D_{2}k_{\perp}^{2}$, $ M(k) = M_{0} + B_{1}k_{z}^{2} + B_{2}k_{\perp}^{2}$ and $ k_{\pm} = k_{x} \pm ik_{y}$.

In the above Hamiltonian, the Fermi-level $ \mu $ has been set to zero. Also, the pair potential $ \Delta_{1} $ is assumed for the zero-momentum fermion pair $ \psi_{1\uparrow}(k_{\parallel},z)\psi_{1\downarrow}(-k_{\parallel},z) $. For the corresponding, $ \psi_{1\downarrow}(k_{\parallel},z)\psi_{1\uparrow}(-k_{\parallel},z) $, the pair potential is $ -\Delta_{1} $.~\cite{heikkila2013physics,mackenzie2003superconductivity} 

A more accurate order parameter can be obtained through a self-consistent calculation. For instance the order parameter $ \Delta_{1} $, is given as 
\begin{equation}
\Delta_{1} = f(z)\int dk_{\parallel}\langle\psi_{1\uparrow}(k_{\parallel},z)\psi_{1\downarrow}(-k_{\parallel},z)
\label{delta1}
\end{equation}
In eq.~\ref{delta1}, $\psi_{1\uparrow}$ and $\psi_{1\downarrow}$ refer to the wave function components in the 8 $\times $ 1 column vector that correspond to $ \vert P1_{z}^{+} \uparrow \rangle $ and $ -\vert P1_{z}^{+} \downarrow \rangle $. Details of the self-consistent calculation of the order parameter is discussed in Section~\ref{res}.

As a concrete demonstration of band-gap alteration, an \textit{s}-wave superconductor which is characterized by an isotropic energy-gap and conveniently represented as $ \Delta = i\Delta_{0}\sigma_{y} $ is considered. The surface states can be modeled through a Dirac Hamiltonian written as $ H_{0}(k) = v_{f}(\sigma_{x}k_{x} + \sigma_{y}k_{y})-\mu $. Adding a pair potential such that fermions with opposite spins and momentum are coupled, the Hamiltonian takes a simple form
\begin{equation}
\begin{pmatrix}
H_{0}(k) & \Delta(k) \\
-\Delta^{*}(-k) & -H_{0}(-k) \\
\end{pmatrix}
\label{red_ham}
\end{equation}
For an s-type superconductor, the pair potential is independent of the $ \overrightarrow{k} $, the Hamiltonian given in Eq.(~\ref{red_ham}) can now be solved to obtain a dispersion relationship of the form 
\begin{equation}
\varepsilon = \pm\sqrt{(v_{f}|\textbf{k}| \pm\mu)^{2} + \Delta_{0}^{2}}
\label{sti}
\end{equation}
For a p-type superconductor, the pair potential is generally expressed as $ \Delta(k) = k_{x} \pm ik_{y} $. The corresponding dispersion gives the following eigenvalues
\begin{equation}
\varepsilon = \pm v_{f}|\textbf{k}|\pm\Delta_{k}
\label{pti}
\end{equation} 
It is easy to see that at the $ \Gamma $ point the pair potential is equal to zero and band gap closing is preserved. This is qualitatively different from the \textit{s}-wave case discussed earlier.~\cite{linder2010interplay}

\subsection{Applied Magnetic field}
Applied magnetic field represents a significant way to open a band gap in the spectrum near $ k = 0 $. The $ k = 0 $ degeneracy is protected by time reversal symmetry which can be destroyed by a magnetic field. The applied magnetic field is incorporated in the Hamiltonian through the standard Peierls substitution $ k = k -\dfrac{e}{c}A $ where $ \overrightarrow{A} $ is the magnetic vector potential. The simplest way to add a magnetic field is layer the surface of a TI with a ferromagnet. The ferromagnetic proximity exchange field polarized along an arbitrary direction~\cite{rachel2014giant} can be written in the most general way as $ \overrightarrow{m_{1}} = m_{1}(sin\theta cos\phi, sin\theta sin\phi, cos\theta) $. To illustrate the effect of a magnetic field, the spectrum of a TI slab under a $\overrightarrow{B}$ field directed parallel and perpendicular to the surface is computed. For a parallel orientation, the states remain gap-less but are shifted along the in-plane direction perpendicular to the field. The surface states around each Dirac cone are approximately described as 
\begin{equation}
H_{xy} = v_{f}(\sigma_{x}k_{y} - \sigma_{y}k_{x})
\label{simple_dirac}
\end{equation}
This gives a linear dispersion relationship of the form $\varepsilon = \pm v_{f}\sqrt{(k_{x}^{2} + k_{y}^{2}})$. A $ \overrightarrow{B} $ field perpendicular to the surface (0,0,$\overrightarrow{B}$) can be represented using a Landau gauge $ \overrightarrow{A} = (-\overrightarrow{B}y,0,0) $. Using the standard Peierls transformation, the simple Dirac Hamiltonian is transformed to 
\begin{equation}
H_{xy} = v_{f}\begin{pmatrix}
0 & k_{y} + i(k_{x} + eBy \\
k_{y} - i(k_{x} +eBy) & 0 \\
\end{pmatrix}
\end{equation}
The eigen values are Landau levels given as $ \varepsilon = \pm v_{f}\sqrt{2neB} $. A simpler Dirac Hamiltonian which captures the phenomenlogical behaviour of surface electrons in presence of the exchange field is shown in Eq.~\ref{dferro}.
\begin{equation}
H = \hbar v_{f}k\cdot\sigma + m\cdot\sigma
\label{dferro}
\end{equation}

However, for a surface parallel to the magnetic field, the same gauge can now be written as $ \overrightarrow{A} = (-Bz,0,0) $. Using the same Peierls transformation, the $ \overrightarrow{k}_{x} $ vector is now transformed to $ \overrightarrow{k_{x}} = \overrightarrow{k}_{x} + eBz $ . The $ \overrightarrow{k}_{y} $ vector remains unchanged. For an x-y surface, \textit{z} is a constant which means that $ eBz $ is a constant too and commutes with $ \overrightarrow{k}_{x} $. The new set of eigen values are therefore written as 
\begin{equation}
\varepsilon = \pm v_{f}\sqrt{(k_{x} + eBz^{2}) + k_{y}^{2}}
\label{bfield_par} 
\end{equation}
The only difference is that Dirac cones now appear at $ ( \overrightarrow{k_{x}} = -eBz, \overrightarrow{k_{y}} = 0 ) $. In other words, the eigen energies of the system are independent of the choice of gauge. 

\section{Results}
\label{res}
\subsection{Comparison between surface states in Bi$_{2}$Se$_{3}$ and edge states in HgTe}
Two structures are considered to understand the geometric dependence of surface and edge states. 3D topological insulators with surface states are modeled as thin films of finite thickness along the \textit{z} direction while 2D topological insulators exhibiting edge states are nano-ribbons with two dimensional confinement along \textit{y} and \textit{z} directions. It must be pointed out here that a nano-ribbon is constructed out of an inverted quantum well~\cite{sengupta2013design} to allow topological edge states. 2D TIs are modeled using the BHZ Hamiltonian described in Section~\ref{meta}. For a Bi$_{2}$Se$_{3}$ film, it is observed that surface states are preserved for thickness as low as 9.0 $\mathrm{nm} $ while the corresponding edge states in a CdTe/HgTe/CdTe 2D-TI nanoribbon requires a thickness of at least 100 $ \mathrm{nm} $.
\begin{figure}[h]
\includegraphics[scale=1]{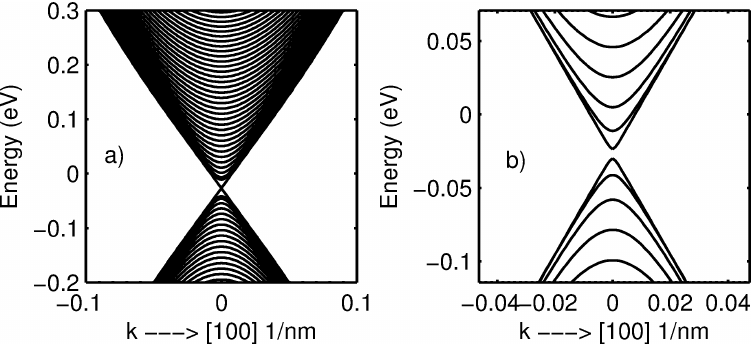}
\caption{Dispersion of a CdTe/HgTe/CdTe nanoribbon construced from an inverted CdTe/HgTe/CdTe quantum well. Fig.~\ref{fig3}a shows a nanoribbon 100.0 $\mathrm{nm}$ wide and 8.0 $\mathrm{nm}$ thick. This structure shows a Dirac crossing and topological edge state. Fig.~\ref{fig3}b is 50.0 $\mathrm{nm}$ wide and 8.0 $\mathrm{nm}$ thick ribbon with a finite band gap.}
\label{fig3}
\end{figure} 
This suggests that much smaller samples that retain their conducting surface states can be fabricated with 3D TIs than in the case of a CdTe/HgTe/CdTe 2D-TI. The physical interpretation of these length scales can be sought in the delocalization length or penetration depth of these states. The penetration depth is usually expressed as $\xi = \hbar v_{f}/M $, where \textit{M} is the bulk band gap and $v_{f}$ is the Fermi-velocity. By just noting the charge excitation gap, it is obvious that large difference in penetration depth can be easily explained. The bulk excitation gap of Bi$_{2}$Se$_{3}$ is much higher than that of HgTe, which is a zero band gap compound.  

\begin{figure}[h]
\includegraphics[scale=0.7]{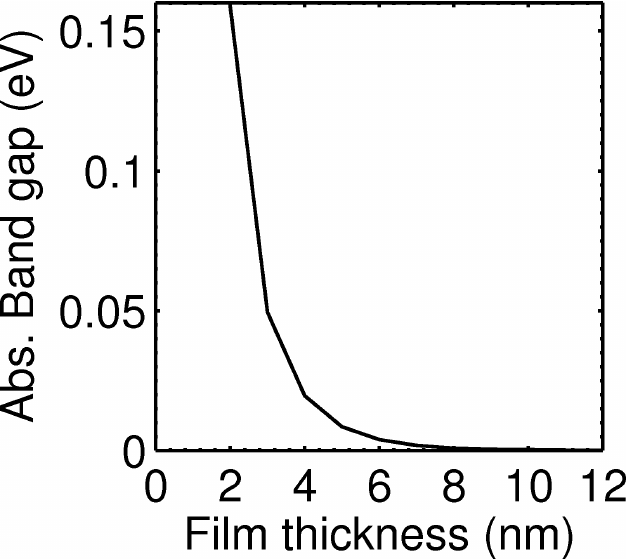}
\caption{Band-gap opening as a function of Bi$_{2}$Se$_{3}$ film thickness. Band gap opens because the two surfaces hybridize.}
\label{fig4}
\end{figure}

Another result that supports this observation is the film thickness at which a gap opens up in a Sb$_{2}$Te$_{3}$ compared to Bi$_{2}$Se$_{3}$. As a numerical example, the band gap opening in a 5.0 $\mathrm{nm}$ thin-film is 0.0084 $\mathrm{eV}$ and 0.0629 $\mathrm{eV}$ in Bi$_{2}$Se$_{3}$ and Sb$_{2}$Te$_{3}$ respectively. Assuming comparable Fermi velocities, Sb$_{2}$Te$_{3}$ which has a smaller bulk band-gap at $\Gamma$, the surface states would be more delocalized than Bi$_{2}$Se$_{3}$. As a result of greater delocalization, complete hybridization occurs and gives a larger band gap opening. The dependence of band-gap for various Bi$_{2}$Se$_{3}$ film-thickness is plotted in Fig.~\ref{fig4}.

\subsection{Asymmetric thin films of 3D-TIs}
\begin{figure}[h]
\includegraphics[scale=0.23]{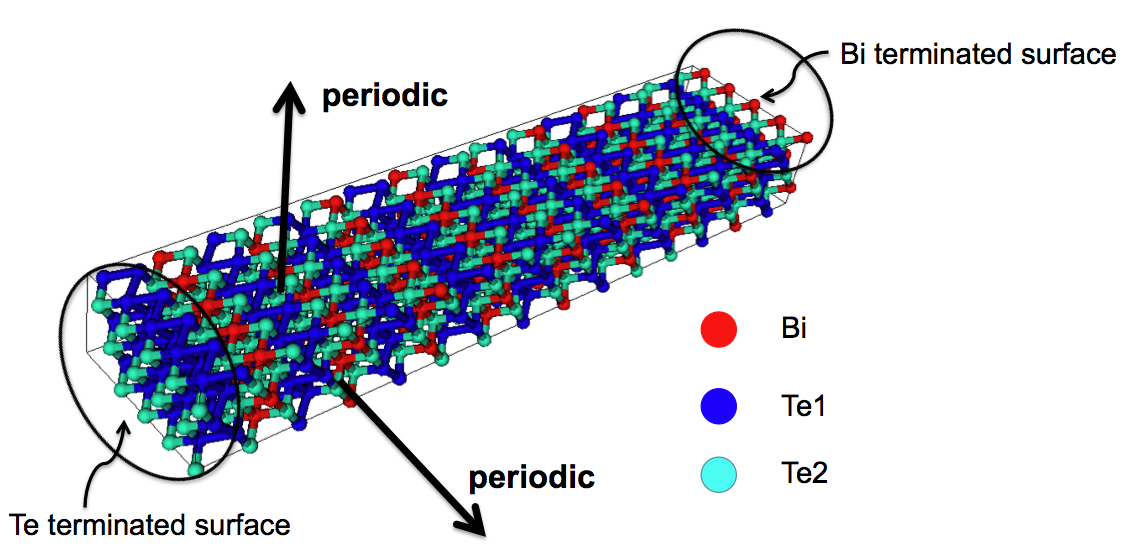}
\caption{A Bi$_{2}$Te$_{3}$ thin film with two different surfaces. The two surfaces have Bi and Te termination thus making them chemically inequivalent}
\label{fig5}
\end{figure}

\begin{figure}[h]
\includegraphics[scale=0.8]{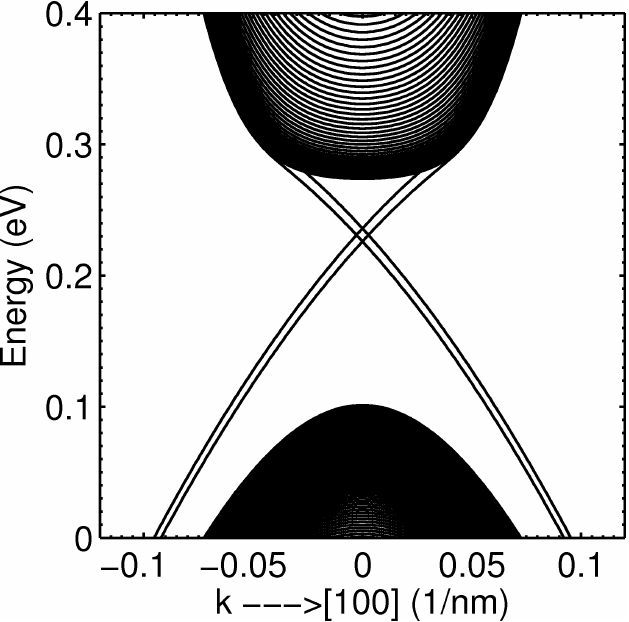}
\caption{A Bi$_{2}$Se$_{3}$ thin film with built-in asymmetry. Asymmetry in this film was artificially introduced by using a small potential along the confinement direction. The two Dirac cones are now separated in energy}
\label{fig6}
\end{figure}

For a symmetric thin film with identical surfaces, the two Dirac cones representing the surface states are degenerate. Asymmetry due to two different surfaces though can exist on account of inequivalent surface termination or presence of a substrate. This naturally occurring asymmetry~\cite{hong2010ultrathin} can be reproduced by applying an electric field along the confinement direction. An example of a real arrangement of atoms in a Bi$_{2}$Te$_{3}$ thin film with Bi and tellurium surface termination is shown in Fig.~\ref{fig5}. Under ideal conditions (ignoring impurity effects), a dipole is formed between the two surfaces which then host a single Dirac cone as shown in Fig.~\ref{fig6}. The separation of the Dirac cones on each surface depends on the asymmetry induced by the potential. 

\begin{figure}[h]
\includegraphics[scale=0.8]{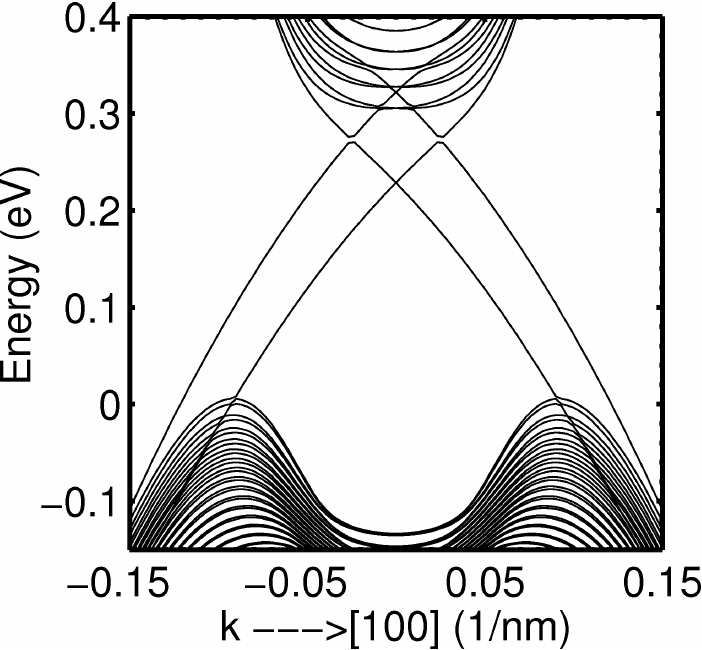}
\caption{An ultra-thin Bi$_{2}$Se$_{3}$ film with asymmetry. The two Dirac hyperbolas from the ultra-thin film in presence of asymmetry are now spin-split. They form four copies, two from conduction and valence band and maintain degeneracy only at the $\Gamma$ point}
\label{fig7}
\end{figure}

A more interesting dispersion relationship arises in the presence of an asymmetric potential in an ultra-thin film. This situation can be broken in to two parts and results obtained above can then be combined. For an ultra-thin film whose surfaces have hybridized, two massive Dirac hyperbolas are created. Each hyperbola is spin-degenerate as shown in Fig.~\ref{fig1}b. If film asymmetry impresses an electrostatic potential breaking inversion symmetry, the two Dirac hyperbolas will now split in to four sets of Dirac hyperbolas (Fig.~\ref{fig7}). The spin degeneracy of the Dirac-hyperbolas is only maintained at the $\Gamma$ because it happens to be one of the time-reversal-invariant-momenta (TRIM) points.~\cite{teo2008surface,essin2007topological}

\begin{figure}[h]
\includegraphics[scale=0.8]{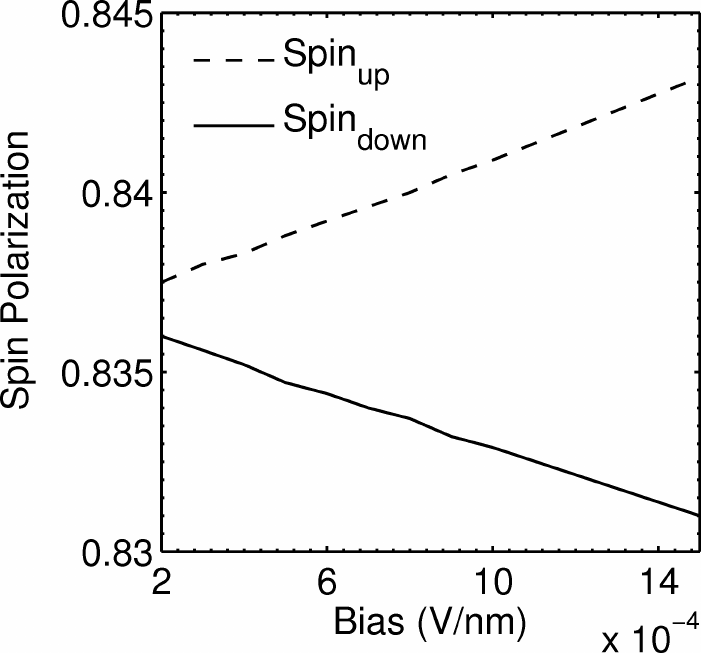}
\caption{Spin-up and spin-down components are of unequal strength in an asymmetric thin film. The difference in amplitude between the two components increase with higher field/asymmetry.}
\label{fig8}
\end{figure}

For a symmetric free-standing film, the spin-up and spin-down components of the in-plane spin-polarization are exactly anti-parallel. Interestingly, if spin-polarization is measured in an asymmetric 3D-TI film, the two anti-parallel spin components are no longer of same strength. The amplitude of each spin-component is now a function of the asymmetry expressed as an electric potential. The two surfaces of a 3D-TI therefore have unequal spin-amplitudes. Fig.~\ref{fig8} was obtained by computing the spin polarization for $ \overrightarrow{k_{x}} = 0.01 nm^{-1} $. The \textit{k}-vector is chosen close to the $ \Gamma $ point to satisfactorily represent the dispersion in the Dirac regime. 

\subsection{Superconductor and TI heterostructure}
\begin{figure}[h]
\includegraphics[scale=1]{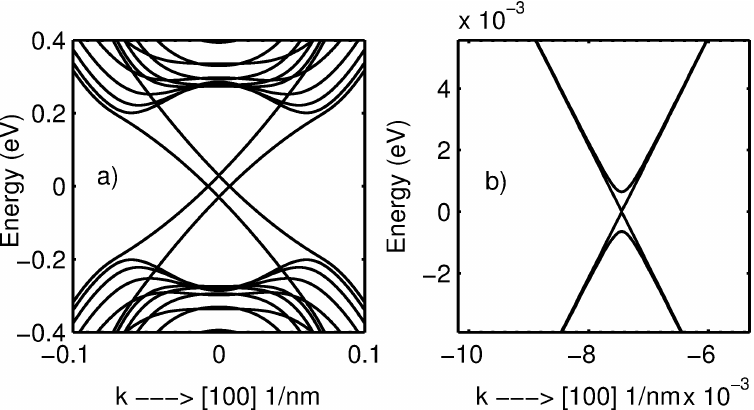}
\caption{The revamped surface dispersion of a 50.0 $ \mathrm{nm} $ Bi$_{2}$Se$_{3}$ film when coated with an \textit{s}-wave superconductor. Fig.~\ref{fig9}a shows the overall band dispersion while Fig.~\ref{fig9}b displays the energy dispersion around the Dirac cones. The surface with no superconductor penetration has a TI surface state.}
\label{fig9}
\end{figure}

The surface band dispersion of a 3D TI is first presented followed by a self-consistent calculation of the order parameter. A 50.0 $ \mathrm{nm} $ Bi$_{2}$Se$_{3}$ film was layered with an \textit{s}-wave superconductor. The superconductor is assumed to extend up to 25.0 $ \mathrm{nm} $. The remaining half of the slab is pristine Bi$_{2}$Se$_{3}$ and possesses regular 3D TI properties. The \textit{s}-wave superconductor is assumed to have the material properties of Aluminium~\cite{fossheim2005superconductivity} whose order parameter($\Delta_{1}$ in Eq.~\ref{bdg_full}) at \textit{T} = 0 K is roughly equal to 0.34 $ \mathrm{meV} $. Band dispersion of the surface states for the Bi$_{2}$Se$_{3}$ film coupled to the superconductor is shown in Fig.~\ref{fig9}a. Since the superconductor extends only until half of the structure, the second surface still shows a Dirac cone while the top surface has an open band gap. This is shown in the zoomed out Fig.~\ref{fig9}b. Additionally, in this calculation, the chemical potential $ \mu $ and the second order parameters $ \Delta_{2}$ has been set to zero.

To compute the spatial profile of the order parameter self-consistently, $ \Delta_{1}$ is initially set to the bulk superconducting gap value for Aluminium. With this value, the TI-SC Hamiltonian($H_{TS}$)is diagonalized and a new order parameter $ \Delta_{1_{n}}$ using Eq.~\ref{delta1}is produced. In Eq.~\ref{delta1}, the function $ f(z) $ is set equal to the bulk superconducting energy gap. $ \Delta_{1}$ in the TI-SC Hamiltonian($H_{TS}$) is set to $ \Delta_{1_{n}}$ and the process is iterated until convergence is achieved. The spatial profile shows that $ \Delta_{1}$ is significantly suppressed at 25.0 $ \mathrm{nm} $ mark which is the TI-SC interface and drops to zero beyond that as expected since the superconductor extends only until the halfway mark. The interface therefore strongly distorts the superconductor wave function as demonstrated through this self-consistent calculation. The oscillations seen in the order parameter is due to finite discretization along the quantized \textit{z}-direction. For calculations presented here, a matrix of the size 2008 $\times $ 2008 was diagonalized. Two spatial profiles are included for $ \mu $ = 0 and $ \mu $ = 0.2 $\mathrm{eV}$. As evident, the chemical potential does not significantly impact the order parameter. 
\begin{figure}
\includegraphics[scale=1]{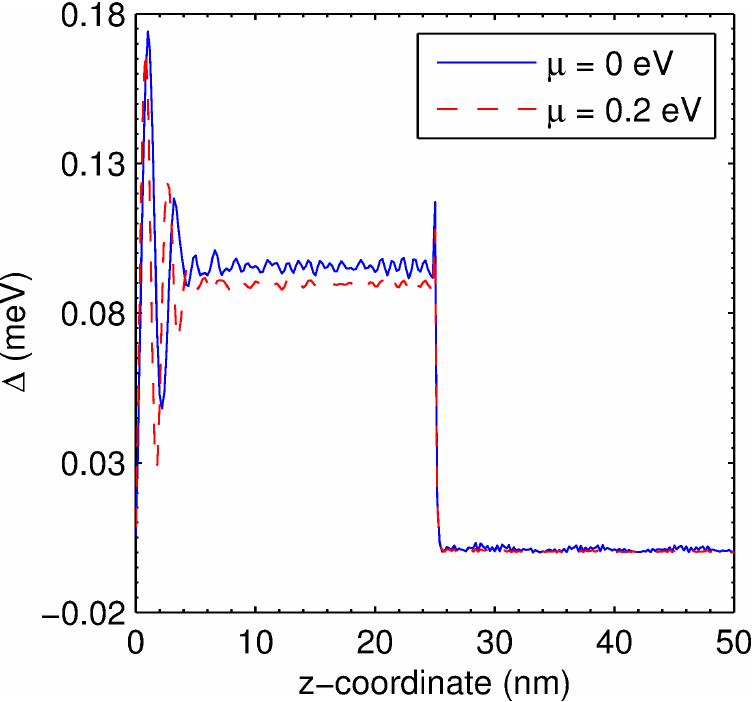}
\caption{Superconducting order parameter $ \Delta_{1}$ calculated self-consistently at two different chemical potentials is plotted as a function of the quantized \textit{z}-axis. The Bi$_{2}$Se$_{3}$ film is 50.0 $ \mathrm{nm} $ in thickness and superconductor is assumed to  extend until 25.0 $\mathrm{nm}$.}
\label{fig10}
\end{figure}

\subsection{Spectrum of TI slabs in presence of an exchange field}
\begin{figure}[tb]
\includegraphics[scale=1]{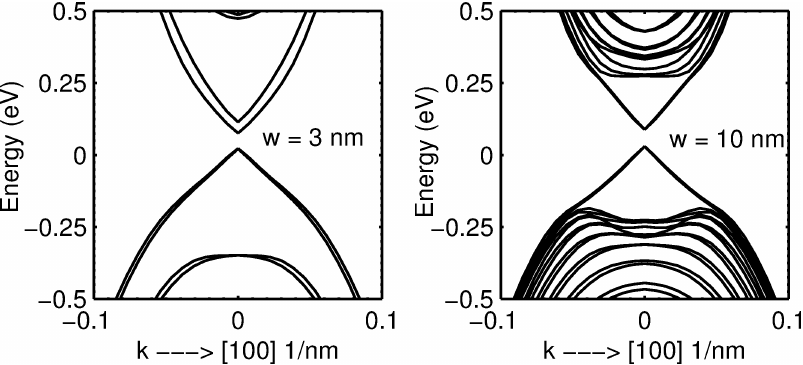}
\caption{Dispersion plots for two TI slabs in presence of a 30.0 $\mathrm{meV}$ \textit{z}-directed exchange field. The exchange field gives different band gap openings for a 3.0 $ \mathrm{nm} $ film (left) with hybridized surfaces in contrast to decoupled surfaces of a thicker 10.0 $ \mathrm{nm} $ film.}
\label{fig11}
\end{figure}

An exchange field operational in a ferromagnet and with a component along the normal (the normal is chosen to be along the \textit{z}-axis)to the TI film surface opens a band gap. A relatively thick slab of Bi$_{2}$Se$_{3}$ (10.0 $ \mathrm{nm} $ in present calculation) which ensures a decoupled top and bottom surface in presence of a ferromagnetic exchange field ($ \Delta_{z}$) shows a band gap opening. A gap equal to twice the exchange energy appears and the bands acquire a parabolic character. The linear dispersion, though is preserved away from the $ \Gamma $ point and the perpendicular spin-momentum locking, is observed again. For thin slabs which allow hybridization of the two surfaces, the Dirac bands change to Dirac hyperbolas as mentioned elsewhere in this paper. In presence of an exchange field, which breaks time reversal symmetry, and near \textit{k} = 0 the bands split in two states with spin up energies $ \pm E_{g}/2 + \Delta_{z} $ and two spin down states $ \pm E_{g}/2 - \Delta_{z} $. For an exchange energy $ \delta_{z} = 30 meV $, the gap induced in a thick TI slab as shown in Fig.~\ref{fig11}a is equal to 0.0585 $\mathrm{eV}$. This split is roughly twice the chosen exchange energy. The corresponding split (Fig.~\ref{fig11}b) for a thin slab (3.0 $ \mathrm{nm} $) is computed to be 0.0510 $\mathrm{eV}$. 

\begin{figure}[t]
\includegraphics[scale=0.9]{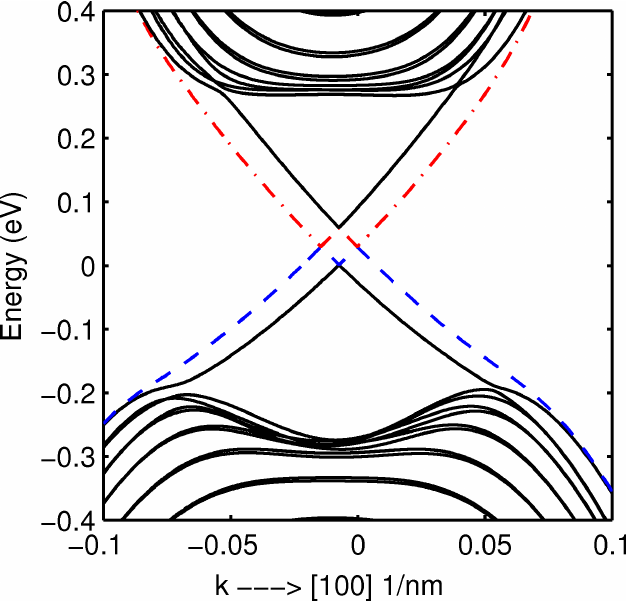}
\caption{Dispersion plot for a 10 $\mathrm{eV}$ TI slabs with an in-plane magnetic field (along \textit{y}-axis) of 10.0 Tesla. The Dirac cone shifted along \textit{k$_{x}$}-axis comes from the ferromagnet coated surface.}
\label{fig12}
\end{figure}

Finally, as a demonstration of the assertion made and analytically derived in Eq.(~\ref{bfield_par}), the axis of magnetization for the ferromagnet is assumed to lie along the \textit{y}-axis. The magnetic field equal to 10.0 Tesla, in Landau gauge is represented as $ \overrightarrow{A} = \left(B_{y}z, 0, 0 \right) $ where \textit{z} is the set of discretized coordinates along the confined axis. This choice of magnetic vector potential can be easily verified through the standard $ \overrightarrow{B} $ = $\nabla \times \vec{A} $ relation to represent a $ \overrightarrow{B} $ field along the \textit{y}-axis. Further, it was also assumed that the ferromagnetic proximity effect extends only for the  first three layers of the film. Two Dirac cones are formed as shown in Fig.~\ref{fig12}. One of the Dirac cones which corresponds to the surface unaffected by the ferromagnet is at the $\Gamma $ point. 
The second Dirac cone which is from the top surface coated with a ferromagnet has shifted along the \textit{k$_{x}$}-axis to -0.01403 nm$^{-1}$. The energies of the two Dirac cones are equal to 0.02914 $ \mathrm{eV} $. The valence and conduction band component are clearly identified in blue and red colour respectively  in Fig.~\ref{fig12}. It is therefore imperative to have an out-of-plane magnetization for a finite band gap and destruction of topological states.  

\section{Conclusions} 
\label{conc}
The results presented above demonstrate the finite-sized effects on 2D and 3D topological insulator states. 3D TI films of thickness as low as 6.0 $ \mathrm{nm} $ were found to support surface states. In contrast, a 2D TI CdTe/HgTe/CdTe nanoribbon must be atleast 100.0 $ \mathrm{nm} $ in thickness to prevent hybridization of edge states and open a band gap. The role of asymmetry in a 3D-TI was found to gproduce Dirac hyperbolas which for a hybridized thin film is split similar to a Rashba splitting. It was also shown that influence of an external magnetic field can be emulated by layering the surface of a 3D TI with a ferromagnet. The important condition to note is that the axis of magnetization must have a component perpendicular to the surface. A ferromagnet whose magnetic axis lies in the plane of the surface will shift the Dirac cone and no band gap opening is observed. Finally, coating a 3D-TI with an \textit{s}-wave superconductor leads to a band gap on account of the well-known proximity effect. The band-gap opening depends on the strength of the order parameter which was computed self-consistently in the calculations shown.

\begin{acknowledgements}
One of us (PS) wishes to thank Chandan Setty from the Dept. of Physics at Purdue University for several illuminating discussions on superconductors. We also acknowledge support from Intel Corp. during the early stages of this work.
\end{acknowledgements}

\bibliographystyle{apsrev}

\begin{thebibliography}{35}
\expandafter\ifx\csname natexlab\endcsname\relax\def\natexlab#1{#1}\fi
\expandafter\ifx\csname bibnamefont\endcsname\relax
  \def\bibnamefont#1{#1}\fi
\expandafter\ifx\csname bibfnamefont\endcsname\relax
  \def\bibfnamefont#1{#1}\fi
\expandafter\ifx\csname citenamefont\endcsname\relax
  \def\citenamefont#1{#1}\fi
\expandafter\ifx\csname url\endcsname\relax
  \def\url#1{\texttt{#1}}\fi
\expandafter\ifx\csname urlprefix\endcsname\relax\def\urlprefix{URL }\fi
\providecommand{\bibinfo}[2]{#2}
\providecommand{\eprint}[2][]{\url{#2}}

\bibitem[{\citenamefont{Hasan and Kane}(2010)}]{hasanrmp}
\bibinfo{author}{\bibfnamefont{M.~Z.} \bibnamefont{Hasan}} \bibnamefont{and}
  \bibinfo{author}{\bibfnamefont{C.~L.} \bibnamefont{Kane}},
  \bibinfo{journal}{Rev. Mod. Phys.} \textbf{\bibinfo{volume}{82}},
  \bibinfo{pages}{3045} (\bibinfo{year}{2010}).

\bibitem[{\citenamefont{Fu and Kane}(2007)}]{fuprb07}
\bibinfo{author}{\bibfnamefont{L.}~\bibnamefont{Fu}} \bibnamefont{and}
  \bibinfo{author}{\bibfnamefont{C.~L.} \bibnamefont{Kane}},
  \bibinfo{journal}{Phys. Rev. B} \textbf{\bibinfo{volume}{76}},
  \bibinfo{pages}{045302} (\bibinfo{year}{2007}).

\bibitem[{\citenamefont{Qi and Zhang}(2011)}]{qi2011topological}
\bibinfo{author}{\bibfnamefont{X.-L.} \bibnamefont{Qi}} \bibnamefont{and}
  \bibinfo{author}{\bibfnamefont{S.-C.} \bibnamefont{Zhang}},
  \bibinfo{journal}{Reviews of Modern Physics} \textbf{\bibinfo{volume}{83}},
  \bibinfo{pages}{1057} (\bibinfo{year}{2011}).

\bibitem[{\citenamefont{Kane and Mele}(2005)}]{kaneqsh05}
\bibinfo{author}{\bibfnamefont{C.~L.} \bibnamefont{Kane}} \bibnamefont{and}
  \bibinfo{author}{\bibfnamefont{E.~J.} \bibnamefont{Mele}},
  \bibinfo{journal}{Phys. Rev. Lett.} \textbf{\bibinfo{volume}{95}},
  \bibinfo{pages}{146802} (\bibinfo{year}{2005}).

\bibitem[{\citenamefont{Murakami}(2007)}]{murakami07}
\bibinfo{author}{\bibfnamefont{S.}~\bibnamefont{Murakami}},
  \bibinfo{journal}{New Journal of Physics} \textbf{\bibinfo{volume}{9}},
  \bibinfo{pages}{356} (\bibinfo{year}{2007}).

\bibitem[{\citenamefont{Xia et~al.}(2009)\citenamefont{Xia, Qian, Hsieh, Wray,
  Pal, Lin, Bansil, Grauer, Hor, Cava et~al.}}]{xia09obsv}
\bibinfo{author}{\bibfnamefont{Y.}~\bibnamefont{Xia}},
  \bibinfo{author}{\bibfnamefont{D.}~\bibnamefont{Qian}},
  \bibinfo{author}{\bibfnamefont{D.}~\bibnamefont{Hsieh}},
  \bibinfo{author}{\bibfnamefont{L.}~\bibnamefont{Wray}},
  \bibinfo{author}{\bibfnamefont{A.}~\bibnamefont{Pal}},
  \bibinfo{author}{\bibfnamefont{H.}~\bibnamefont{Lin}},
  \bibinfo{author}{\bibfnamefont{A.}~\bibnamefont{Bansil}},
  \bibinfo{author}{\bibfnamefont{D.}~\bibnamefont{Grauer}},
  \bibinfo{author}{\bibfnamefont{Y.}~\bibnamefont{Hor}},
  \bibinfo{author}{\bibfnamefont{R.}~\bibnamefont{Cava}}, \bibnamefont{et~al.},
  \bibinfo{journal}{Nature Physics} \textbf{\bibinfo{volume}{5}},
  \bibinfo{pages}{398} (\bibinfo{year}{2009}).

\bibitem[{\citenamefont{Chen et~al.}(2009)\citenamefont{Chen, Analytis, Chu,
  Liu, Mo, Qi, Zhang, Lu, Dai, Fang et~al.}}]{chen09exp}
\bibinfo{author}{\bibfnamefont{Y.}~\bibnamefont{Chen}},
  \bibinfo{author}{\bibfnamefont{J.}~\bibnamefont{Analytis}},
  \bibinfo{author}{\bibfnamefont{J.}~\bibnamefont{Chu}},
  \bibinfo{author}{\bibfnamefont{Z.}~\bibnamefont{Liu}},
  \bibinfo{author}{\bibfnamefont{S.}~\bibnamefont{Mo}},
  \bibinfo{author}{\bibfnamefont{X.}~\bibnamefont{Qi}},
  \bibinfo{author}{\bibfnamefont{H.}~\bibnamefont{Zhang}},
  \bibinfo{author}{\bibfnamefont{D.}~\bibnamefont{Lu}},
  \bibinfo{author}{\bibfnamefont{X.}~\bibnamefont{Dai}},
  \bibinfo{author}{\bibfnamefont{Z.}~\bibnamefont{Fang}}, \bibnamefont{et~al.},
  \bibinfo{journal}{Science} \textbf{\bibinfo{volume}{325}},
  \bibinfo{pages}{178} (\bibinfo{year}{2009}).

\bibitem[{\citenamefont{Roy}(2009)}]{roy09}
\bibinfo{author}{\bibfnamefont{R.}~\bibnamefont{Roy}}, \bibinfo{journal}{Phys.
  Rev. B} \textbf{\bibinfo{volume}{79}}, \bibinfo{pages}{195322}
  (\bibinfo{year}{2009}).

\bibitem[{\citenamefont{Carpentier}(2014)}]{carpentier2014topology}
\bibinfo{author}{\bibfnamefont{D.}~\bibnamefont{Carpentier}},
  \bibinfo{journal}{arXiv preprint arXiv:1408.1867}  (\bibinfo{year}{2014}).

\bibitem[{\citenamefont{Franz}(2013)}]{franz2013topological}
\bibinfo{author}{\bibfnamefont{M.}~\bibnamefont{Franz}},
  \emph{\bibinfo{title}{Topological Insulators}} (\bibinfo{publisher}{Elsevier
  Science}, \bibinfo{address}{Burlington}, \bibinfo{year}{2013}), ISBN
  \bibinfo{isbn}{0444633146}.

\bibitem[{\citenamefont{Dresselhaus et~al.}(2008)\citenamefont{Dresselhaus,
  Dresselhaus, and Jorio}}]{mildred2008group}
\bibinfo{author}{\bibfnamefont{M.~S.} \bibnamefont{Dresselhaus}},
  \bibinfo{author}{\bibfnamefont{G.}~\bibnamefont{Dresselhaus}},
  \bibnamefont{and} \bibinfo{author}{\bibfnamefont{A.}~\bibnamefont{Jorio}},
  \emph{\bibinfo{title}{Group Theory}} (\bibinfo{publisher}{Springer},
  \bibinfo{year}{2008}).

\bibitem[{\citenamefont{Zhang et~al.}(2009)\citenamefont{Zhang, Liu, Qi, Dai,
  Fang, and Zhang}}]{zhang09}
\bibinfo{author}{\bibfnamefont{H.}~\bibnamefont{Zhang}},
  \bibinfo{author}{\bibfnamefont{C.}~\bibnamefont{Liu}},
  \bibinfo{author}{\bibfnamefont{X.}~\bibnamefont{Qi}},
  \bibinfo{author}{\bibfnamefont{X.}~\bibnamefont{Dai}},
  \bibinfo{author}{\bibfnamefont{Z.}~\bibnamefont{Fang}}, \bibnamefont{and}
  \bibinfo{author}{\bibfnamefont{S.}~\bibnamefont{Zhang}},
  \bibinfo{journal}{Nature Physics} \textbf{\bibinfo{volume}{5}},
  \bibinfo{pages}{438} (\bibinfo{year}{2009}).

\bibitem[{\citenamefont{Fu}(2009)}]{fu2009hexagonal}
\bibinfo{author}{\bibfnamefont{L.}~\bibnamefont{Fu}},
  \bibinfo{journal}{Physical review letters} \textbf{\bibinfo{volume}{103}},
  \bibinfo{pages}{266801} (\bibinfo{year}{2009}).

\bibitem[{\citenamefont{Liu et~al.}(2010)\citenamefont{Liu, Qi, Zhang, Dai,
  Fang, and Zhang}}]{liu2010model}
\bibinfo{author}{\bibfnamefont{C.-X.} \bibnamefont{Liu}},
  \bibinfo{author}{\bibfnamefont{X.-L.} \bibnamefont{Qi}},
  \bibinfo{author}{\bibfnamefont{H.}~\bibnamefont{Zhang}},
  \bibinfo{author}{\bibfnamefont{X.}~\bibnamefont{Dai}},
  \bibinfo{author}{\bibfnamefont{Z.}~\bibnamefont{Fang}}, \bibnamefont{and}
  \bibinfo{author}{\bibfnamefont{S.-C.} \bibnamefont{Zhang}},
  \bibinfo{journal}{Physical Review B} \textbf{\bibinfo{volume}{82}},
  \bibinfo{pages}{045122} (\bibinfo{year}{2010}).

\bibitem[{\citenamefont{Sakamoto et~al.}(2010)\citenamefont{Sakamoto, Hirahara,
  Miyazaki, Kimura, and Hasegawa}}]{sakamoto2010spectroscopic}
\bibinfo{author}{\bibfnamefont{Y.}~\bibnamefont{Sakamoto}},
  \bibinfo{author}{\bibfnamefont{T.}~\bibnamefont{Hirahara}},
  \bibinfo{author}{\bibfnamefont{H.}~\bibnamefont{Miyazaki}},
  \bibinfo{author}{\bibfnamefont{S.-i.} \bibnamefont{Kimura}},
  \bibnamefont{and} \bibinfo{author}{\bibfnamefont{S.}~\bibnamefont{Hasegawa}},
  \bibinfo{journal}{Physical Review B} \textbf{\bibinfo{volume}{81}},
  \bibinfo{pages}{165432} (\bibinfo{year}{2010}).

\bibitem[{\citenamefont{Chang et~al.}(2010)\citenamefont{Chang, He, Liu, Zhang,
  Chen, Wang, Ma, Wang, and Xue}}]{chang2010growth}
\bibinfo{author}{\bibfnamefont{C.-Z.} \bibnamefont{Chang}},
  \bibinfo{author}{\bibfnamefont{K.}~\bibnamefont{He}},
  \bibinfo{author}{\bibfnamefont{M.-H.} \bibnamefont{Liu}},
  \bibinfo{author}{\bibfnamefont{Z.-C.} \bibnamefont{Zhang}},
  \bibinfo{author}{\bibfnamefont{X.}~\bibnamefont{Chen}},
  \bibinfo{author}{\bibfnamefont{L.-L.} \bibnamefont{Wang}},
  \bibinfo{author}{\bibfnamefont{X.-C.} \bibnamefont{Ma}},
  \bibinfo{author}{\bibfnamefont{Y.-Y.} \bibnamefont{Wang}}, \bibnamefont{and}
  \bibinfo{author}{\bibfnamefont{Q.-K.} \bibnamefont{Xue}},
  \bibinfo{journal}{arXiv preprint arXiv:1012.5716}  (\bibinfo{year}{2010}).

\bibitem[{\citenamefont{Bychkov and Rashba}(1984)}]{bychkov1984properties}
\bibinfo{author}{\bibfnamefont{Y.~A.} \bibnamefont{Bychkov}} \bibnamefont{and}
  \bibinfo{author}{\bibfnamefont{E.}~\bibnamefont{Rashba}},
  \bibinfo{journal}{JETP lett} \textbf{\bibinfo{volume}{39}},
  \bibinfo{pages}{78} (\bibinfo{year}{1984}).

\bibitem[{\citenamefont{Kuemmeth and Rashba}(2009)}]{kuemmeth2009giant}
\bibinfo{author}{\bibfnamefont{F.}~\bibnamefont{Kuemmeth}} \bibnamefont{and}
  \bibinfo{author}{\bibfnamefont{E.~I.} \bibnamefont{Rashba}},
  \bibinfo{journal}{Physical Review B} \textbf{\bibinfo{volume}{80}},
  \bibinfo{pages}{241409} (\bibinfo{year}{2009}).

\bibitem[{\citenamefont{e~Silva et~al.}(1997)\citenamefont{e~Silva, La~Rocca,
  and Bassani}}]{e1997spin}
\bibinfo{author}{\bibfnamefont{E.~d.~A.} \bibnamefont{e~Silva}},
  \bibinfo{author}{\bibfnamefont{G.}~\bibnamefont{La~Rocca}}, \bibnamefont{and}
  \bibinfo{author}{\bibfnamefont{F.}~\bibnamefont{Bassani}},
  \bibinfo{journal}{Physical Review B} \textbf{\bibinfo{volume}{55}},
  \bibinfo{pages}{16293} (\bibinfo{year}{1997}).

\bibitem[{\citenamefont{Sengupta et~al.}(2013)\citenamefont{Sengupta, Kubis,
  Tan, Povolotskyi, and Klimeck}}]{sengupta2013design}
\bibinfo{author}{\bibfnamefont{P.}~\bibnamefont{Sengupta}},
  \bibinfo{author}{\bibfnamefont{T.}~\bibnamefont{Kubis}},
  \bibinfo{author}{\bibfnamefont{Y.}~\bibnamefont{Tan}},
  \bibinfo{author}{\bibfnamefont{M.}~\bibnamefont{Povolotskyi}},
  \bibnamefont{and} \bibinfo{author}{\bibfnamefont{G.}~\bibnamefont{Klimeck}},
  \bibinfo{journal}{Journal of Applied Physics} \textbf{\bibinfo{volume}{114}},
  \bibinfo{pages}{043702} (\bibinfo{year}{2013}).

\bibitem[{\citenamefont{Bernevig}(2013)}]{bernevig2013topological}
\bibinfo{author}{\bibfnamefont{B.~A.} \bibnamefont{Bernevig}},
  \emph{\bibinfo{title}{Topological Insulators and Topological
  Superconductors}} (\bibinfo{publisher}{Princeton University Press},
  \bibinfo{year}{2013}).

\bibitem[{\citenamefont{Rothe et~al.}(2010)\citenamefont{Rothe, Reinthaler,
  Liu, Molenkamp, Zhang, and Hankiewicz}}]{rothe2010fingerprint}
\bibinfo{author}{\bibfnamefont{D.}~\bibnamefont{Rothe}},
  \bibinfo{author}{\bibfnamefont{R.}~\bibnamefont{Reinthaler}},
  \bibinfo{author}{\bibfnamefont{C.}~\bibnamefont{Liu}},
  \bibinfo{author}{\bibfnamefont{L.}~\bibnamefont{Molenkamp}},
  \bibinfo{author}{\bibfnamefont{S.}~\bibnamefont{Zhang}}, \bibnamefont{and}
  \bibinfo{author}{\bibfnamefont{E.}~\bibnamefont{Hankiewicz}},
  \bibinfo{journal}{New Journal of Physics} \textbf{\bibinfo{volume}{12}},
  \bibinfo{pages}{065012} (\bibinfo{year}{2010}).

\bibitem[{\citenamefont{Tinkham}(2012)}]{tinkham2012introduction}
\bibinfo{author}{\bibfnamefont{M.}~\bibnamefont{Tinkham}},
  \emph{\bibinfo{title}{Introduction to superconductivity}}
  (\bibinfo{publisher}{Courier Dover Publications}, \bibinfo{year}{2012}).

\bibitem[{\citenamefont{De~Gennes}(1999)}]{de1999superconductivity}
\bibinfo{author}{\bibfnamefont{P.}~\bibnamefont{De~Gennes}},
  \emph{\bibinfo{title}{Superconductivity of Metals and Alloys (Advanced Book
  Classics)}} (\bibinfo{publisher}{Addison-Wesley Publ. Company Inc},
  \bibinfo{year}{1999}).

\bibitem[{\citenamefont{Fetter and Walecka}(2003)}]{fetter2003quantum}
\bibinfo{author}{\bibfnamefont{A.~L.} \bibnamefont{Fetter}} \bibnamefont{and}
  \bibinfo{author}{\bibfnamefont{J.~D.} \bibnamefont{Walecka}},
  \emph{\bibinfo{title}{Quantum theory of many-particle systems}}
  (\bibinfo{publisher}{Courier Dover Publications}, \bibinfo{year}{2003}).

\bibitem[{\citenamefont{Poole et~al.}(1999)\citenamefont{Poole, Farach, and
  Creswick}}]{poole1999handbook}
\bibinfo{author}{\bibfnamefont{C.~K.} \bibnamefont{Poole}},
  \bibinfo{author}{\bibfnamefont{H.~A.} \bibnamefont{Farach}},
  \bibnamefont{and} \bibinfo{author}{\bibfnamefont{R.~J.}
  \bibnamefont{Creswick}}, \emph{\bibinfo{title}{Handbook of
  superconductivity}} (\bibinfo{publisher}{Academic Press},
  \bibinfo{year}{1999}).

\bibitem[{\citenamefont{Orlando and Delin}(1991)}]{orlando1991foundations}
\bibinfo{author}{\bibfnamefont{T.~P.} \bibnamefont{Orlando}} \bibnamefont{and}
  \bibinfo{author}{\bibfnamefont{K.~A.} \bibnamefont{Delin}},
  \emph{\bibinfo{title}{Foundations of applied superconductivity}}, vol.
  \bibinfo{volume}{224} (\bibinfo{publisher}{Addison-Wesley Reading, MA},
  \bibinfo{year}{1991}).

\bibitem[{\citenamefont{Heikkil{\"a}}(2013)}]{heikkila2013physics}
\bibinfo{author}{\bibfnamefont{T.~T.} \bibnamefont{Heikkil{\"a}}},
  \emph{\bibinfo{title}{The Physics of Nanoelectronics: Transport and
  Fluctuation Phenomena at Low Temperatures}}, vol.~\bibinfo{volume}{21}
  (\bibinfo{publisher}{Oxford University Press}, \bibinfo{year}{2013}).

\bibitem[{\citenamefont{Mackenzie and
  Maeno}(2003)}]{mackenzie2003superconductivity}
\bibinfo{author}{\bibfnamefont{A.~P.} \bibnamefont{Mackenzie}}
  \bibnamefont{and} \bibinfo{author}{\bibfnamefont{Y.}~\bibnamefont{Maeno}},
  \bibinfo{journal}{Reviews of Modern Physics} \textbf{\bibinfo{volume}{75}},
  \bibinfo{pages}{657} (\bibinfo{year}{2003}).

\bibitem[{\citenamefont{Linder et~al.}(2010)\citenamefont{Linder, Tanaka,
  Yokoyama, Sudb{\o}, and Nagaosa}}]{linder2010interplay}
\bibinfo{author}{\bibfnamefont{J.}~\bibnamefont{Linder}},
  \bibinfo{author}{\bibfnamefont{Y.}~\bibnamefont{Tanaka}},
  \bibinfo{author}{\bibfnamefont{T.}~\bibnamefont{Yokoyama}},
  \bibinfo{author}{\bibfnamefont{A.}~\bibnamefont{Sudb{\o}}}, \bibnamefont{and}
  \bibinfo{author}{\bibfnamefont{N.}~\bibnamefont{Nagaosa}},
  \bibinfo{journal}{Physical Review B} \textbf{\bibinfo{volume}{81}},
  \bibinfo{pages}{184525} (\bibinfo{year}{2010}).

\bibitem[{\citenamefont{Rachel and Ezawa}(2014)}]{rachel2014giant}
\bibinfo{author}{\bibfnamefont{S.}~\bibnamefont{Rachel}} \bibnamefont{and}
  \bibinfo{author}{\bibfnamefont{M.}~\bibnamefont{Ezawa}},
  \bibinfo{journal}{Physical Review B} \textbf{\bibinfo{volume}{89}},
  \bibinfo{pages}{195303} (\bibinfo{year}{2014}).

\bibitem[{\citenamefont{Hong et~al.}(2010)\citenamefont{Hong, Kundhikanjana,
  Cha, Lai, Kong, Meister, Kelly, Shen, and Cui}}]{hong2010ultrathin}
\bibinfo{author}{\bibfnamefont{S.~S.} \bibnamefont{Hong}},
  \bibinfo{author}{\bibfnamefont{W.}~\bibnamefont{Kundhikanjana}},
  \bibinfo{author}{\bibfnamefont{J.~J.} \bibnamefont{Cha}},
  \bibinfo{author}{\bibfnamefont{K.}~\bibnamefont{Lai}},
  \bibinfo{author}{\bibfnamefont{D.}~\bibnamefont{Kong}},
  \bibinfo{author}{\bibfnamefont{S.}~\bibnamefont{Meister}},
  \bibinfo{author}{\bibfnamefont{M.~A.} \bibnamefont{Kelly}},
  \bibinfo{author}{\bibfnamefont{Z.-X.} \bibnamefont{Shen}}, \bibnamefont{and}
  \bibinfo{author}{\bibfnamefont{Y.}~\bibnamefont{Cui}}, \bibinfo{journal}{Nano
  letters} \textbf{\bibinfo{volume}{10}}, \bibinfo{pages}{3118}
  (\bibinfo{year}{2010}).

\bibitem[{\citenamefont{Teo et~al.}(2008)\citenamefont{Teo, Fu, and
  Kane}}]{teo2008surface}
\bibinfo{author}{\bibfnamefont{J.~C.} \bibnamefont{Teo}},
  \bibinfo{author}{\bibfnamefont{L.}~\bibnamefont{Fu}}, \bibnamefont{and}
  \bibinfo{author}{\bibfnamefont{C.}~\bibnamefont{Kane}},
  \bibinfo{journal}{Physical Review B} \textbf{\bibinfo{volume}{78}},
  \bibinfo{pages}{045426} (\bibinfo{year}{2008}).

\bibitem[{\citenamefont{Essin and Moore}(2007)}]{essin2007topological}
\bibinfo{author}{\bibfnamefont{A.~M.} \bibnamefont{Essin}} \bibnamefont{and}
  \bibinfo{author}{\bibfnamefont{J.}~\bibnamefont{Moore}},
  \bibinfo{journal}{Physical Review B} \textbf{\bibinfo{volume}{76}},
  \bibinfo{pages}{165307} (\bibinfo{year}{2007}).

\bibitem[{\citenamefont{Fossheim and
  Sudb{\o}}(2005)}]{fossheim2005superconductivity}
\bibinfo{author}{\bibfnamefont{K.}~\bibnamefont{Fossheim}} \bibnamefont{and}
  \bibinfo{author}{\bibfnamefont{A.}~\bibnamefont{Sudb{\o}}},
  \emph{\bibinfo{title}{Superconductivity: physics and applications}}
  (\bibinfo{publisher}{John Wiley \& Sons}, \bibinfo{year}{2005}).

\end{thebibliography}

\end{document}